\documentclass[sigconf]{acmart}
\usepackage{xcolor}

\newif\ifshowrevisions
\ifdefined\SHOWREVISIONS
  \showrevisionstrue
\else
  \showrevisionsfalse
\fi

\definecolor{reviewdelete}{RGB}{180,0,0}
\definecolor{reviewadd}{RGB}{0,70,170}
\ifshowrevisions
  \DeclareRobustCommand{\reviewdel}[1]{\textcolor{reviewdelete}{[DEL: #1]}}
  \DeclareRobustCommand{\reviewadd}[1]{\textcolor{reviewadd}{#1}}
\else
  \DeclareRobustCommand{\reviewdel}[1]{}
  \DeclareRobustCommand{\reviewadd}[1]{#1}
\fi

\copyrightyear{2025}
\acmYear{2025}
\setcopyright{rightsretained}
\acmConference[CSCW Companion '25]{Companion of the Computer-Supported Cooperative Work and Social Computing}{October 18--22, 2025}{Bergen, Norway}
\acmBooktitle{Companion of the Computer-Supported Cooperative Work and Social Computing (CSCW Companion '25), October 18--22, 2025, Bergen, Norway}\acmDOI{10.1145/3715070.3749241}
\acmISBN{979-8-4007-1480-1/2025/10}

\AtBeginDocument{%
  \providecommand\BibTeX{{%
    \normalfont B\kern-0.5em{\scshape i\kern-0.25em b}\kern-0.8em\TeX}}}

\begin{document}

\title{Echo Chambers and Information Brokers on Truth Social: A Study of Network Dynamics and Political Discourse}

\author{Emelia May Hughes}
\email{ehughes8@nd.edu}
\affiliation{%
  \institution{University of Notre Dame}
  \state{Indiana}
  \country{USA}
}
\author{Tim Weninger}
\email{tweninger@nd.edu}
\affiliation{%
  \institution{ND-IBM Technology Ethics Lab}
  \institution{University of Notre Dame}
  \state{Indiana}
  \country{USA}
}

\renewcommand{\shortauthors}{Emelia May Hughes \& Tim Weninger}

\begin{abstract}
This study examines the structural dynamics of Truth Social, a politically aligned social media platform, during two major political events: the U.S. Supreme Court’s overturning of \textit{Roe v. Wade} and the FBI’s search of Mar-a-Lago. Using a large-scale dataset of user interactions based on re-truths (platform-native reposts), we analyze how the network evolves in relation to fragmentation, polarization, and user influence. Our findings reveal a segmented and ideologically homogenous structure dominated by a small number of central figures. Political events prompt temporary consolidation around shared narratives, followed by rapid returns to fragmented, echo-chambered clusters. Centrality metrics highlight the disproportionate role of key influencers, particularly @realDonaldTrump, in shaping visibility and directing discourse. These results contribute to research on alternative platforms, political communication, and online network behavior, demonstrating how infrastructure and community dynamics together reinforce ideological boundaries and limit cross-cutting engagement.
\end{abstract}



\maketitle

\ifshowrevisions
  \noindent\reviewdel{Deleted text}\quad\reviewadd{Added text}\par\medskip
\fi

\section{Introduction}

The landscape of online political communication in the United States has grown increasingly fragmented, particularly following the deplatforming of Donald Trump from mainstream social media. This moment catalyzed the creation of Truth Social, a platform intended for Trump supporters and users disillusioned with conventional outlets~\cite{gerard2023truth}. Truth Social \reviewdel{reflects a broader shift toward ideologically aligned digital enclaves, where insular discourse and political polarization are intensified} \reviewadd{provides a case of an ideologically positioned social media environment}~\cite{fuchs2017social,bright2018fragmentation}. Prior research \reviewdel{shows that social media often reinforces in-group identity, as users preferentially engage with content that affirms their political beliefs} \reviewadd{finds that user choices and network composition can reduce exposure to ideologically cross-cutting content}~\cite{bakshy2015exposure,barbera2020social,bright2018fragmentation}. \reviewdel{These tendencies are especially pronounced in echo chambers, where exposure to dissenting views is limited and information cascades can amplify polarizing narratives} \reviewadd{Research on echo chambers connects network homophily and selective exposure to separated patterns of political communication}~\cite{cinelli2021echo,colleoni2014echo}.

\reviewdel{While most studies focus on large-scale networks like Facebook and Twitter, relatively little empirical work has examined smaller, ideologically homogeneous platforms such as Truth Social} \reviewadd{A substantial share of foundational empirical research examines large platforms such as Facebook and Twitter, while empirical study of Truth Social was limited when its dataset was introduced in 2023}~\cite{barbera2020social,gerard2023truth}. \reviewdel{Though often peripheral to broader public discourse, these spaces may act as incubators for political narratives that later migrate outward} \reviewadd{Research on other fringe and alternative communities shows that narratives can travel between smaller communities and larger platforms, although that evidence does not itself establish diffusion from Truth Social}~\cite{zannettou2018origins}. \reviewdel{The presence of influencers and information brokers within these networks suggests that, even in seemingly closed systems, external content occasionally penetrates and shapes discourse in complex ways} \reviewadd{Prior work treats political influence as multidimensional and shows that cross-platform influence can be estimated from observed diffusion patterns}~\cite{dubois2014influence,zannettou2018origins,gerard2023truth}.

\begin{figure}
\centering
  \begin{minipage}[c]{0.45\textwidth}
    \includegraphics[width = \textwidth]{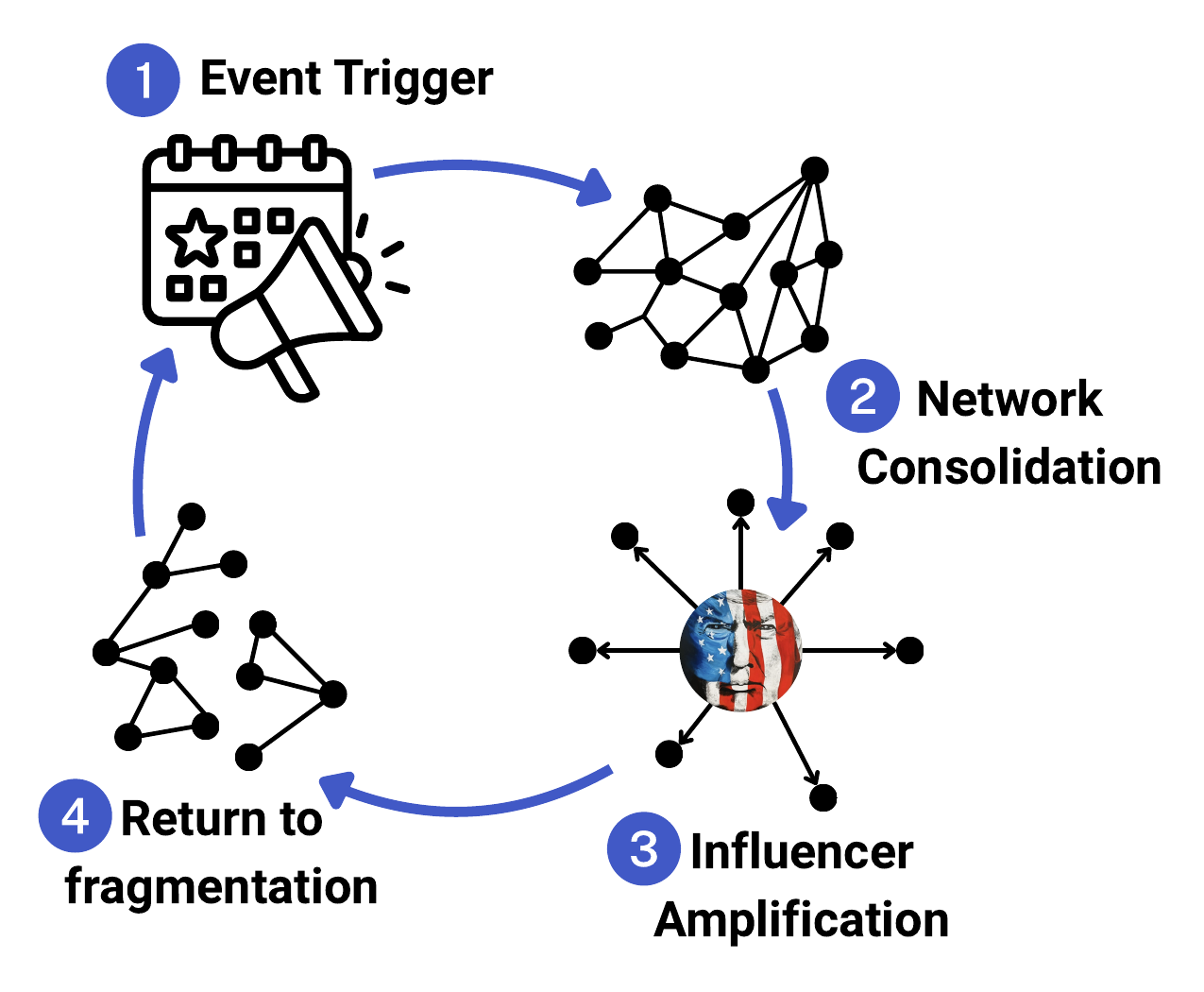}
  \end{minipage}
  \begin{minipage}[c]{0.45\textwidth}
    \caption{Cyclical response model of Truth Social’s network dynamics during major political events. High-salience moments trigger temporary consolidation and heightened influencer activity, followed by a return to fragmented, ideologically homogeneous communities. The cycle reflects how structural insularity and centralized amplification co-produce polarized engagement patterns.}
    \label{fig:influence-cycle}
  \end{minipage}
  \hfill
\end{figure}

In this context, Truth Social offers a valuable lens into how platform design, user alignment, and structural fragmentation intersect. \reviewdel{Building on prior research into networked political engagement} \reviewadd{Building on research connecting network structure, ideological alignment, and platform-specific communication patterns to political fragmentation}~\cite{bright2018fragmentation,cinelli2021echo,yarchi2021polarization}, we examine how the platform’s network structure responds to two major events: the overturning of Roe v. Wade and the FBI search of Mar-a-Lago (Trump's property). These moments serve as entry points for understanding both stability and disruption within an ideologically concentrated space.

We pose the following research question: How does the structure of the Truth Social network shift around key political events?


We hypothesize that Truth Social exhibits a high degree of homophily, driven by ideological self-selection and limited external linkage. Using user-level attributes, such as in-degree (re-Truths received) and out-degree (re-Truths made), alongside network-level measures like clustering coefficient and average path length, we assess the cohesion and segmentation of the platform. While narrative analysis could deepen understanding of discursive content, this study focuses on structural properties as a foundation for future work on information flow and ideological reinforcement.

\paragraph{Findings in Brief.}
Our analysis reveals a highly clustered, polarized network, with users organizing into ideologically homogeneous communities. Influential accounts---particularly @realDonaldTrump—dominate the flow of attention, with centrality metrics spiking during major events. Both the Roe v. Wade decision and the FBI’s Mar-a-Lago search produced short-term network consolidation, followed by a return to fragmented echo chambers. These patterns underscore the platform’s responsiveness to political events while affirming its long-term structural insularity and the central role of influencers in shaping internal discourse.

This work contributes to ongoing CSCW discussions of sociotechnical systems by revealing how platform infrastructure, political alignment, and user behavior interact to reinforce ideological boundaries. By introducing a cyclical model of network response to political events, our study also offers insight into how emergent attention dynamics on alternative platforms may influence broader media ecosystems. These findings have implications for platform governance, algorithmic design, and moderation strategies aimed at mitigating polarization.

\section{Methodology}
\subsection{Data Collection}

We analyze data from Truth Social spanning its launch on February 21, 2022, through October 15, 2022. The platform does not offer a public API, so we used a custom web scraper to extract publicly visible user activity from the site’s web interface. This process adhered to platform policies and did not violate its \texttt{robots.txt} file or terms of service \cite{gerard2023truth}.

The resulting dataset includes 823{,}927 posts (“Truths”) from 454{,}458 users---approximately 20\% of registered accounts---and captures the full posting history of the 65{,}536 most active users. Data collection began with @realDonaldTrump and expanded through a breadth-first traversal of follower and following connections, producing a structurally meaningful sample of user interactions and platform activity.

\subsection{Limitations}

Although the dataset is extensive, several limitations apply:

\begin{itemize}
    \item \textbf{Truncated follower data:} Only the first 50 followers of any user were accessible, which prevented full reconstruction of follower networks for high-profile accounts.
    \item \textbf{Request rate constraints:} While the platform does not enforce explicit rate limits, practical constraints such as request timing and server response latency shaped the scraping process.
    \item \textbf{Sampling bias:} Seeding the crawl with @realDonaldTrump likely introduced bias toward politically active users. Even so, the network’s short average path lengths and the variety of content types suggest that the dataset captures a broad cross-section of platform activity.
\end{itemize}

\subsection{Ethical Considerations}

All data were collected from publicly accessible content. No private or login-restricted information was accessed at any stage. Users on Truth Social post with the expectation of public visibility. This study was reviewed by the University of Notre Dame’s institutional review board and deemed exempt from full human subjects review.

\section{Results}
\subsection{Network Structure and User Roles}

This analysis investigates how information circulates on Truth Social by modeling the re-truth network, where edges represent reposted content during two key political events: the U.S. Supreme Court’s overturning of \textit{Roe v. Wade} and the FBI’s search of Mar-a-Lago. The resulting network is highly fragmented but features a centralized core shaped by a small group of highly active users.

\subsubsection{Centrality and Influence}

Across all centrality measures, @realDonaldTrump consistently emerges as the dominant figure within the network. Other high-engagement users, such as @NavyCMC, @WarriorNurse, and @rkblues, fluctuate in prominence depending on the surrounding political context.

\begin{itemize}
    \item \textbf{Degree centrality} highlights the outsized role of @realDonaldTrump and a small group of others in driving re-truth activity.
    \item \textbf{Betweenness centrality} identifies bridging users, including @Sunnyc45 and @butler, who link otherwise distant regions of the network.
    \item \textbf{Closeness centrality} ranks users like @AngieCutlip and @ASoftstar highly, indicating their potential to disseminate information efficiently across communities.
    \item \textbf{Eigenvector centrality} reinforces the influence of @realDonaldTrump and shows that users such as @catturd2 and @DineshDSouza gain prominence through their connections to other highly central accounts.
\end{itemize}

\subsubsection{Fragmentation and Community Structure}

Despite the visibility of a few central figures, the broader network remains sparse and decentralized, with low density ($3.67 \times 10^{-4}$) and numerous strongly connected components. Community detection using the Louvain algorithm reveals 40 distinct clusters, with a modularity score of 0.2828, indicating moderate segmentation. Political events appear to heighten this fragmentation while simultaneously focusing attention on key accounts, amplifying their visibility and influence.

\subsection{Event Response: Overturning of \textit{Roe v. Wade}}

\begin{figure}[t]
    \includegraphics[width=0.9\linewidth]{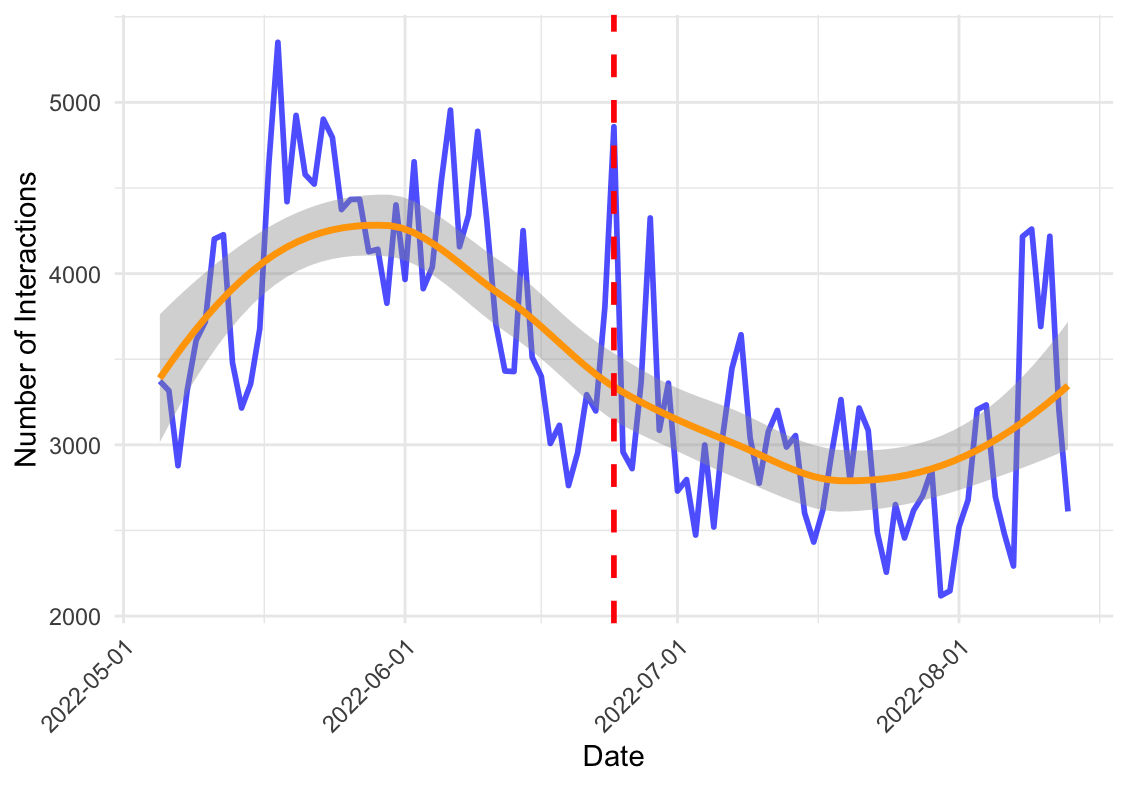}
    \caption{Engagement Trends Around \textit{Roe v. Wade} Overturning}
    \label{fig:roe}
\end{figure}

The Supreme Court’s decision to overturn \textit{Roe v. Wade} produced measurable shifts in Truth Social’s network structure and user engagement. Prior to the decision, the network contained 12{,}621 communities with a high modularity score of 0.9910, reflecting substantial ideological segmentation. As the event unfolded, the number of communities dropped to 5{,}433 and modularity declined slightly to 0.9804, suggesting temporary consolidation around shared narratives.

This centralization did not persist. In the days following the decision, the network rapidly fragmented: 89{,}486 communities emerged, and modularity rose to 0.9991. These shifts signal a return to ideological insularity, as users coalesced into smaller, more homogeneous groups—a pattern consistent with echo chamber reinforcement.

Centrality metrics followed a similar trend. @realDonaldTrump’s degree centrality increased from 3{,}742 to 17{,}409, reflecting a surge in engagement. Betweenness and eigenvector centrality also spiked for several key users, highlighting their elevated influence during periods of political upheaval.

\subsection{Event Response: FBI Raid on Mar-a-Lago}

The FBI’s search of Mar-a-Lago on August 8, 2022, prompted a similar network response. The number of communities dropped from 12{,}621 to 4{,}337, while modularity decreased from 0.9910 to 0.9876. These changes reflect a brief period of network cohesion, as users rallied around the event.

Fragmentation resumed shortly afterward. The community count climbed to 79{,}246, and modularity increased to 0.9989—mirroring the post–\textit{Roe v. Wade} pattern. This cyclical dynamic, where major political events induce temporary consolidation followed by retreat into ideological clusters, underscores the platform’s structurally polarized nature.

Engagement metrics reflect these shifts. Degree centrality for @newsbusters rose from 3{,}125 to 12{,}930, while users like @dbongino gained betweenness centrality, positioning them as key discourse brokers. These shifts illustrate how political flashpoints not only reshape the network structurally but also elevate specific users to prominence.

\begin{figure}[t]
    \includegraphics[width=0.9\linewidth]{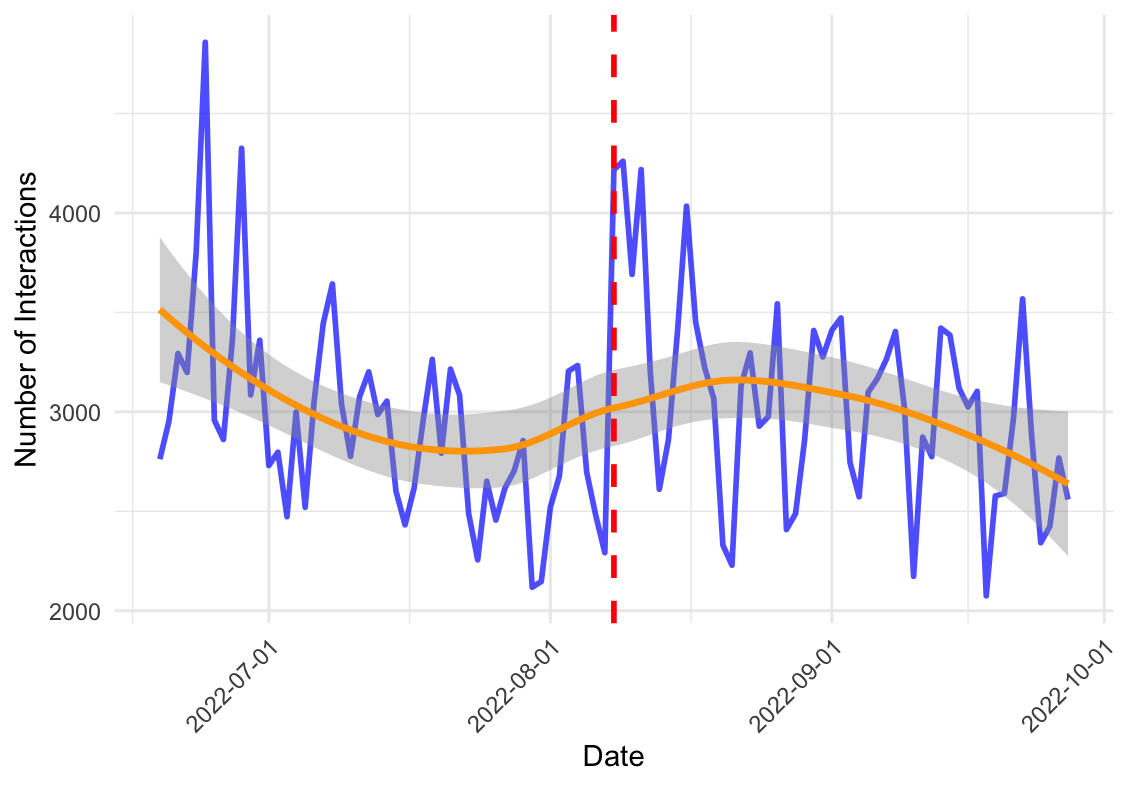}
    \caption{Engagement Trends Around FBI Search of Mar-a-Lago}
    \label{fig:fbi}
\end{figure}

\section{Discussion}
\subsection{Interpreting Network Dynamics}

Truth Social exhibits a highly polarized and ideologically clustered structure. The re-truth mechanism, functioning as an explicit form of endorsement, reinforces segmentation by amplifying aligned narratives rather than encouraging cross-ideological engagement~\cite{metaxas2015retweets}. Users largely organize into cohesive, homogeneous communities that operate as ideological echo chambers.

Influencers and information brokers shape discourse most visibly during high-salience events. Spikes in betweenness and eigenvector centrality during the \textit{Roe v. Wade} decision and Mar-a-Lago raid show how key users steer attention and coordinate messaging. These actors connect disparate parts of the network while consolidating ideological authority, often reinforcing dominant narratives within already-aligned communities.

\subsection{Comparison with Mainstream Platforms}

\reviewdel{Compared to broader platforms like Twitter, Truth Social displays a more insular and ideologically concentrated structure.} \reviewadd{Within the sampled re-truth network, Truth Social displays an insular and ideologically concentrated structure.} While Twitter also exhibits polarization, \reviewdel{its broader user base and algorithmic diversity facilitate more cross-cutting interaction.} \reviewadd{research finds that exposure and political separation vary with network composition, user choice, and platform context}~\cite{bakshy2015exposure,cinelli2021echo,yarchi2021polarization,barbera2020social}. On Truth Social, re-truths act as \reviewdel{strong ideological signals} \reviewadd{sharing signals}, deepening segmentation and reinforcing political identity~\cite{metaxas2015retweets}.

High modularity and recurring fragmentation point to a structural orientation toward partisan discourse. Community formation is event-driven rather than topic-driven, underscoring that Truth Social functions less as a general-purpose platform and more as infrastructure for politically motivated communication.

\subsection{Event-Driven Network Realignment}

Both the \textit{Roe v. Wade} decision and the FBI’s search of Mar-a-Lago triggered temporary network consolidation, followed by renewed fragmentation. However, the scope and longevity of these responses differed.

The \textit{Roe v. Wade} decision prompted sustained engagement and longer-term structural change. Communities reorganized around reproductive rights, and modularity shifts suggest deeper realignment. In contrast, the FBI raid produced a brief spike in centralization, quickly followed by a return to prior silos. While both events mobilized users, only Roe v. Wade appeared to catalyze enduring transformation.

These patterns support a cyclical model of event-driven network response: brief unification around high-salience moments, followed by a return to fragmented, ideologically consistent structures. This pattern is illustrated in Figure~\ref{fig:influence-cycle}. Influencers amplify this cycle. During Roe v. Wade, engagement was more distributed across ideological lines; during Mar-a-Lago, conservative influencers dominated discourse. This contrast further illustrates the platform’s partisan orientation and the central role of high-visibility actors in shaping response trajectories.

\subsection{Implications for Platform Governance and Design}
Our findings offer several implications for platform designers, moderators, and policymakers concerned with polarization and information flow on ideologically aligned platforms. First, the cyclical consolidation---fragmentation model highlights how political flashpoints serve as entry points for rapid discourse realignment. Platforms might anticipate such surges in influencer centrality and community coalescence by deploying targeted moderation strategies or transparency mechanisms during these high-salience moments.

Second, the structural prominence of a small number of users---especially during consolidation phases---suggests that visibility algorithms play a critical role in amplifying particular narratives. Evidence from Twitter shows that ranking can amplify political content, although the present study does not test Truth Social's ranking system~\cite{huszar2022amplification}. Platform architects might consider interventions that promote exposure to diverse voices or limit the automatic elevation of repeat amplifiers to reduce echo chamber reinforcement.

Finally, the structural insularity of Truth Social underscores the challenge of designing for cross-cutting engagement in politically homogeneous environments. Rather than attempting to engineer deliberation directly, platform governance could focus on transparency, content labeling, or friction-based design changes that make the boundaries of ideological clusters more legible to users.

\subsection{Limitations and Future Directions}

This data set captures roughly 20\% of Truth Social’s user base and overrepresents highly active accounts, potentially understating the activity of less engaged or moderate users. Limited access to follower data and private interactions also restricts full network reconstruction.

The present analysis focuses on structural signals, specifically re-truths, which reflect patterns of sharing but do not establish post content or user intent~\cite{metaxas2015retweets}. Future research could expand this approach by incorporating content-based methods to examine narrative structure, emotional tone, and intergroup dialogue. For instance, discourse analysis of posts related to the Roe v. Wade decision might reveal mobilization frames or moral appeals, while content from the Mar-a-Lago raid could highlight conspiracy rhetoric or anti-institutional sentiment. Combining structural and semantic analysis would offer a more comprehensive view of how ideology is reinforced, contested, or negotiated within the platform.

Cross-platform comparisons also represent a valuable direction for future work. Analyzing similar events on platforms like Twitter or Facebook could help isolate platform-specific affordances and diffusion patterns, providing a broader perspective on Truth Social’s role within the online political communication landscape.

As alternative platforms continue to influence public discourse,\reviewadd{~\cite{zannettou2018origins}} understanding their network properties remains an important task for researchers focused on polarization, visibility, and the infrastructure of political life online. The consolidation–fragmentation cycle identified in this study offers a useful framework to theorize how surges of attention and influencer amplification contribute to the persistence of insular discourse. This model may apply to other ideologically aligned platforms and could help explain why temporary unification around major events often fails to produce long-term structural cohesion.

\section{Conclusion}
Truth Social offers a unique opportunity to examine the network dynamics of an ideologically concentrated platform. Although the user base is relatively small, the network structure reveals how partisan discourse circulates, consolidates, and fragments in response to political events. Our findings suggest that information flow is heavily shaped by a small number of highly central users, and that major political moments temporarily unify the network before it returns to a fragmented, echo chamber-like state.

\subsubsection{Networked Influence}

Influential users on Truth Social do more than attract engagement---they structure the flow of information across the platform. Spikes in centrality metrics around @realDonaldTrump, @newsbusters, and other prominent accounts underscore how discourse is routed through key figures. These users occupy structurally advantageous positions that allow them to both amplify dominant narratives and shape broader patterns of visibility.

\subsubsection{Structural Insularity}

Despite periodic surges in connectivity, Truth Social remains structurally insular. High modularity scores and the rapid rebound of community fragmentation after major events point to a network that reinforces ideological segmentation. Unlike more heterogeneous platforms, Truth Social facilitates reinforcement over deliberation, cultivating spaces where consensus forms within clusters rather than across them.

\section*{Code Availability}

All code used for the analysis presented in this paper is publicly available at the following link: \\ \href{https://github.com/emeliahughes/truth-social}{https://github.com/emeliahughes/truth-social} \\The repository includes all necessary scripts to replicate the results. The data is sourced from Gerard et al., and is publically available \cite{gerard2023truth}.

\bibliographystyle{ACM-Reference-Format}
\bibliography{biblio}

\end{document}